\documentclass[runningheads]{llncs}
\usepackage[T1]{fontenc}
\usepackage{graphicx}
\usepackage{booktabs}
\usepackage[misc]{ifsym}

\usepackage{mwe}
\usepackage{multirow}
\usepackage{booktabs}
\usepackage{bm}
\usepackage{xcolor}
\usepackage{adjustbox}
\usepackage{url}
\usepackage{array}
\usepackage{amsmath,amssymb}

\begin{document}


\title{Leveraging LLMs and Heterogeneous Knowledge Graphs for Persona-Driven Session-Based Recommendation}
\titlerunning{Persona-Driven SBRS with LLMs and KG}


\author{Muskan Gupta \and
Suraj Thapa  \and
Jyotsana Khatri}

\authorrunning{Gupta et al.}
\institute{TCS Research, India \\ 
\email{\{muskan.gupta4, suraj.39, khatri.jyotsana\}@tcs.com}}

\maketitle              

\begin{abstract}
    Session-based recommendation systems (SBRS) aim to capture user's short-term intent from interaction sequences. However, the common assumption of anonymous sessions limits personalization, particularly under sparse or cold-start conditions. Recent advances in LLM-augmented recommendation have shown that LLMs can generate rich item representations, but modeling user personas with LLMs remains challenging due to anonymous sessions.
    In this work, we propose a persona-driven session-based recommendation framework that explicitly models latent user personas inferred from a heterogeneous knowledge graph (KG) and integrates them into a data-driven recommendation pipeline. Our framework adopts a two-stage architecture consisting of personalized information extraction and personalized information utilization, inspired by recent chain-of-thought recommendation approaches. 
    In the personalized information extraction stage, we construct a heterogeneous KG that integrates time-independent user–item interactions, item–item relations, item–feature associations, and external metadata from DBpedia. We then learn latent user personas in an unsupervised manner using a Heterogeneous Deep Graph Infomax (HDGI) objective over a KG initialized with LLM-derived item embeddings. In the personalized information utilization stage, the learned persona representations together with LLM-derived item embeddings are incorporated into a modified architecture of data-driven SBRS to generate a candidate set of relevant items, followed by reranking using the base sequential model to emphasize short-term session intent. 
    Unlike prior approaches that rely solely on sequence modeling or text-based user representations, our method grounds user persona modeling in structured relational signals derived from a heterogeneous KG. Experiments on Amazon Books and Amazon Movies \& TV demonstrate that our approach consistently improves over sequential models with user embeddings derived using session history.

    
\end{abstract}

\section{Introduction}
Session-based recommendation systems (SBRS) have become a dominant paradigm for capturing short-term user intent through interaction sequences, especially in environments characterized by sparse, unreliable, or non-existent long-term historical data. By emphasizing recent item sequences, SBRS methods capture transient preferences such as immediate goals and contextual interests. However, a fundamental limitation of most existing SBRS approaches is their treatment of sessions as smaller anonymous units, overlooking stable user characteristics that persist across time. This assumption significantly constrains personalization, especially in sparse data, and cold-start scenarios, or when user behavior is distributed across disjoint sessions. Recent work has attempted to enrich SBRS with side information or user representations, yet the majority of approaches remain predominantly sequence-centric \cite{narwariya2025semsr,liu2025improving,zhang2025survey,zhang2026rethinking,liu2023enhancing}. As a result, they struggle to disentangle short-term intent from longer-term preference signals. In practice, user behavior is shaped not only by immediate session context but also by latent personas-relatively stable patterns such as brand affinity, aspect-level preferences, stylistic things, or domain-specific inclinations. Explicitly modeling such personas is essential for robust personalization, but is relatively underexplored in SBRS.
\\
At the same time, advances in large language models (LLMs) have opened new possibilities for leveraging textual item metadata, such as titles, descriptions, and reviews. LLMs excel at capturing nuanced semantics that traditional embeddings often fail to encode. However, despite their strength in representing items, LLMs remain less effective and significantly more expensive when used to infer user representations, especially under data scarcity or noisy behavioral histories. Relying solely on LLMs to model both sides of the recommendation problem introduces computational bottlenecks and limits interpretability.
This highlights a fundamental gap: sequential behavior alone is insufficient, and LLMs alone are insufficient. Effective SBRS requires a means to integrate long‑term user preferences with short‑term session dynamics, and to do so in a way that is efficient, interpretable, and robust across sparse data conditions.

In this work, we propose a persona‑driven session-based recommendation framework that addresses this gap by combining the complementary strengths of heterogeneous knowledge graphs (KGs) and LLM-derived representations. Our key insight is that latent user personas-representing stable, long-term preference patterns can be inferred from a rich heterogeneous KG that encodes multi-relational links between users, items, item attributes, and external knowledge sources such as DBpedia. These structured relationships capture stable patterns such as genre affinity, stylistic preferences, or topical interests, which pure sequence models often overlook. 
To realize this idea, we construct a heterogeneous knowledge graph integrating three types of information: (i) time‑independent user–item interactions, (ii) item–item relational structure, and (iii) item–feature associations drawn from metadata sources including DBpedia. We then learn latent user persona representations using a Heterogeneous Deep Graph Infomax (HDGI) objective, which captures high‑order relational semantics in an unsupervised manner. The KG is initialized with LLM-derived item embeddings extracted from textual metadata such as titles and descriptions.
These two complementary components KG-grounded user personas and LLM-derived item embeddings are integrated within a retrieval architecture (data-driven recommendation system). Persona embeddings enrich the retriever with explicit preference cues, while LLM‑based item representations capture fine‑grained semantics essential for high‑quality candidate generation. To further refine predictions, we incorporate reranking using the base model to provide benefit of short-term intent during.
Our main contribution are as follows:
\begin{itemize}
\item Persona‑driven session-based recommendation: We introduce a persona-driven session-based recommendation framework that enhances personalization even in anonymous or sparse settings.
\item KG‑grounded user persona modeling: We construct a heterogeneous KG and employ an unsupervised method for inferring user personas using a Heterogeneous Deep Graph Infomax objective.
\item Hybrid KG $\times$ LLM Architecture: We integrate heterogeneous knowledge graphs with LLM-derived item embeddings to learn user persona representations, combining structured relational signals from the KG with rich semantic information from textual metadata.
\item Knowledge‑Aware Retrieval: We integrate persona embeddings into a retrieval module 
\end{itemize}
Unlike prior SBRS methods that rely solely on sequential behavior or text-derived user representations, our approach infers user personas from structured relational knowledge graphs and integrates them with LLM-derived item semantics. By unifying heterogeneous knowledge graphs with large language models, it closes the gap between short-term intent modeling and long-term preference understanding, offering a robust solution for modern recommendation environments.
\section{Related Work}
\subsection{Session-based and Sequential Recommendation}
Early works on SBRS relied on Markov chain based models with personalization, exemplified by Factorizing Personalized Markov Chains (FPMC), which unified long-term preference modeling and first-order sequential signals for next-basket recommendation \cite{Rendle2010FPMC}.
With the advancements of deep learning, neural architectures became dominant. GRU4Rec \cite{hidasi2015session} introduced recurrent modeling for anonymous sessions, Caser \cite{Tang2018Caser} used convolutional sequence embeddings, SASRec \cite{Kang2018SASRec} applied causal self-attention to balance long-term semantics with few relevant actions, and BERT4Rec \cite{Sun2019BERT4Rec} adopted a bidirectional transformers with an objective for sequence modeling. 

Several attentive SBRS models further focused on capturing short-term session intent. NARM~\cite{Li2017NARM} utilizes a hybrid encoder with attention and STAMP~\cite{Liu2018STAMP} (short-term attention/memory priority) explicitly emphasize session intent and most recent interactions in a session. Despite their effectiveness, sequence-only models typically rely solely on interaction signals and therefore struggle to capture structured user preferences beyond the observed session window, particularly in sparse or cold-start scenarios.

\subsection{Knowledge-Graph–Enhanced Recommendation}
Session-based recommender systems frequently encounter sparsity issues due to insufficient interaction data. Knowledge graphs (KGs) containing auxiliary information can mitigate this issue \cite{kwon2024reckg}. By modeling multi-relational connections, KGs enrich recommendation systems with structured semantic relations among users, items, and attributes. Furthermore, KG-based approaches mitigate hallucinations and improve generalization and interpretability in LLM-based tasks\cite{agrawal-etal-2024-knowledge}. Recent studies have demonstrated the effectiveness of integrating knowledge graphs into LLM-based recommendation systems \cite{Liang2025,wang2025knowledge,zhang2024side}.The efficacy of such system often stems from established graph-based mechanisms. Representative methods include RippleNet~\cite{Wang2018RippleNet}, which propagates user preferences over entity neighborhoods to capture multi-hop semantics; KGCN~\cite{Wang2019KGCN} samples receptive fields and aggregates neighbors with bias; and KGAT~\cite{Wang2019KGAT} further improves recommendation by employing attentive, high-order embedding propagation on hybrid user–item–attribute graphs. 

While these approaches effectively leverage relational knowledge, they typically integrate knowledge graphs with recommendation signals in a loosely coupled manner, and user representations are still primarily derived from interaction histories. As a result, they do not explicitly learn transferable user personas grounded in knowledge graphs. Our work instead learns KG-grounded user \emph{personas} and injects them into the sequential recommendation backbone, enabling both improved personalization and reasoning.

\subsection{Heterogeneous Graph Representation Learning}
Heterogeneous graph neural networks (HGNNs) explcitly model type-specific nodes, edges, and meta-relations. HAN~\cite{Wang2019HAN} proposes hierarchical attention at node and semantic (meta-path) levels; HGT~\cite{Hu2020HGT} introduces type-dependent attention with scalable heterogeneous sampling suitable for web-scale graphs. For unsupervised node representations, Deep Graph Infomax (DGI) maximizes mutual information between local node patches and global summaries and has inspired heterogeneous extensions for contrastive pretraining \cite{Velickovic2019DGI}. 
Inspired by this line of work, we adopt a heterogeneous DGI~\cite{ren2019heterogeneous} style pretraining over a KG that integrates user–item interactions and DBpedia-derived attributes to yield stable, transferable user personas that complement short-term session intent. 

\subsection{LLM-augmented Recommendation and CoT-style Modeling}
LLMs can be used across different stages of the recommendation pipeline \cite{wu2024survey,lin2025can}. 
LLM-augmented recommendation incorporates large language models to encode item semantics, enrich feature representations. Prompting to augment item text with richer semantic descriptions before feeding them into a recommender model substantially improve downstream performance \cite{lyu-etal-2024-llm}. CoT-Rec~\cite{liu2025improving} introduces a chain-of-thought (CoT) framework that derives both user and item representations through LLM-based reasoning over interaction histories and item descriptions. Surveys~\cite{wu2024survey,lin2025can} discuss the accuracy gains alongside high inference cost and limited structural grounding. However, relying on LLMs to generate both item and user representations introduces significant computational overhead and limited structural grounding. Our approach builds on this line: we retain LLM-derived \emph{item} embeddings 
but \emph{replace} LLM-generated user embeddings with KG-grounded personas learned via heterogeneous contrastive pretraining-preserving item-side semantic strength while improving scalability, grounding, and interpretability and grounding of user modeling. 

\section{Preliminaries}
\label{sec:preliminaries}

In this section, we introduce the fundamental concepts and notation used throughout the paper, including session-based recommendation, heterogeneous knowledge graph, and the representation learning components which serve as building blocks of our framework.

\subsection{Basic Definitions and Notations}
\subsubsection{Session-based Recommendation Systems (SBRS):}
Suppose that $\mathcal{S}$ denotes the set of all sessions containing user-item interactions (e.g. click/view/order), and $\mathcal{I}$ denotes the set of $n$ items observed in $\mathcal{S}$.
Any session $s \in \mathcal{S}$ is an ordered sequence of item interactions generated by an anonymous user within a short time window: 
$s = (i_{s,1},i_{s,2},\ldots,i_{s,|s|})$, where  $i_{s,j}$ ($j=1\ldots |s|$) $\in$ $\mathcal{I}$, denotes the  $j^{th}$  item in session $s$.
The objective of SBRS is to predict the next item $i_{s,|s|+1}$. This is typically formulated as an $n$-way classification problem by estimating the $n$-dimensional item-probability vector $\mathbf{\hat{y}}_{s,|s|+1}$ corresponding to the relevance scores for the $n$ items. The $K$ items with the highest scores constitute the top-$K$ recommendations. 

\subsubsection{Heterogeneous Knowledge Graphs:}
A heterogeneous knowledge graph (HKG) is a multi-relational graph
$\mathcal{G}=(\mathcal{V},\mathcal{E},\phi,\psi)$ in which each node $v\in\mathcal{V}$ and each edge
$e\in\mathcal{E}$ has a type specified by the mappings
$\phi:\mathcal{V}\!\to\!\mathcal{T}_V$ (node types) and $\psi:\mathcal{E}\!\to\!\mathcal{T}_E$ (edge types). 
A graph is considered heterogeneous if it contains multiple node or edge types, i.e.,$|\mathcal{T}_V| + |\mathcal{T}_E| \ge 2$. Nodes may contain attribute information represented by an initial feature matrix: $\mathbf{X}$.
\\
\textit{Heterogeneous Graph Representation Learning}: Given a heterogeneous graph $\mathcal{G}$ and the initial feature matrix $\mathbf{X}$, the representation learning task in $\mathcal{G}$ is to
learn the low dimensional node representations $\mathcal{H} \in \mathcal{R}^{|\mathcal{V}|\times d}$
which contains both structure information from graph $\mathcal{G}$ and semantic information from node
attributes of $\mathbf{X}$. Such embeddings enable downstream tasks such as recommendation.
\subsection{Problem Formulation}
In this work, our goal is to effectively leverage KG‑grounded user personas/preferences and rich item semantics to improve session-based recommendation under the anonymous session assumption. The input consists of a set of anonymous sessions $\mathcal{S}$, a heterogeneous KG $\mathcal{G}=(\mathcal{V}, \mathcal{E}, \phi, \psi)$ which integrates time-independent user-item interactions, item-item relations (e.g., also-buy, also-view), and item-feature associations (e.g. brands, category), and external semantic metadata derived from DBPedia). Using this graph, we infer a KG-grounded user personas that capture latent user preferences. Additionally, we derive semantic item embeddings using large language models (LLMs) from textual signals such as product descriptions and metadata.

\section{Methodology}
\begin{figure}
    \vspace{-60pt}
    \centering
    \includegraphics[width=1\linewidth]{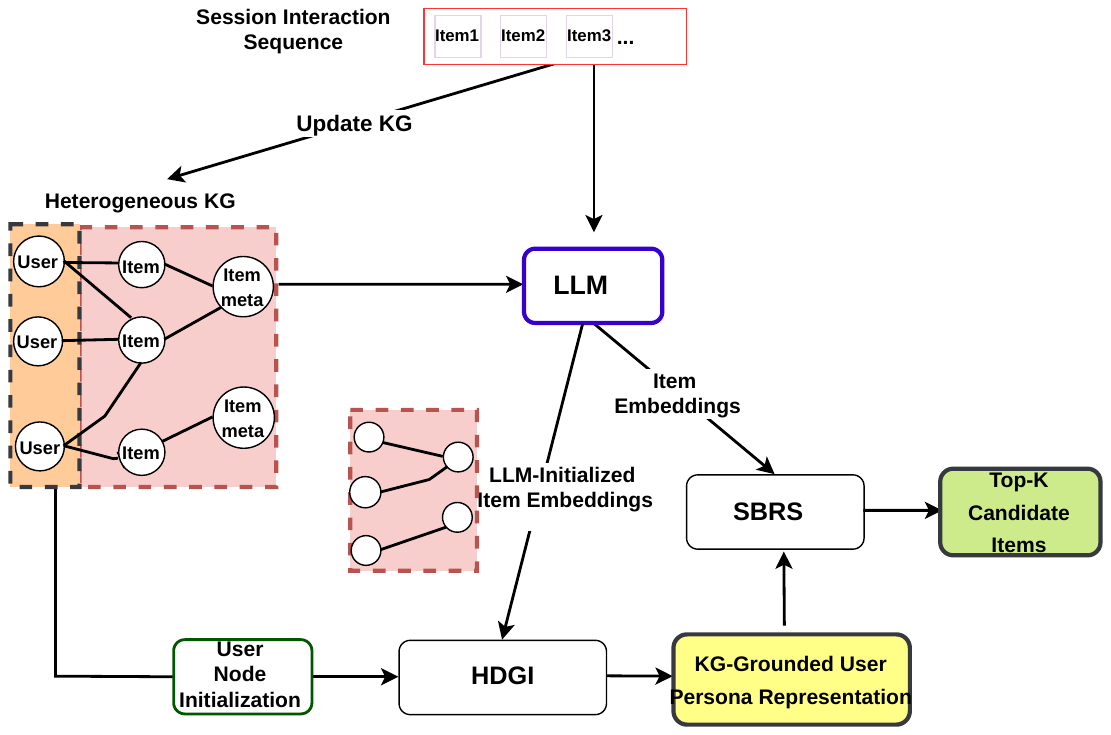}
    \vspace{-220pt}
    \caption{\centering An overview of our proposed framework: Persona-driven session based recommendation via knowledge graph}
    \label{fig:framework}
    \vspace{-10pt}
\end{figure}
Figure \ref{fig:framework} presents an overview of the proposed framework for persona-driven session-based recommendation via knowledge graphs. In this section, we introduce a persona-driven session-based recommendation framework that leverages heterogeneous knowledge graphs (HKGs) to infer stable user preferences beyond short-term interaction signals.
Inspired by CoTRec~\cite{liu2025improving}, our framework consists of two phases,
(i) \emph{Personalized Information Extraction}, which constructs an LLM-initialized heterogeneous knowledge graph (HKG) and learns KG-grounded user personas through unsupervised representation learning and (ii) \emph{Personalized Information Utilization}, which leverages the extracted persona and semantic signals in a data-driven recommendation model and rerank it using a base sequential model.The framework uses KG-driven persona modeling to capture user preferences beyond short-term session signals, while using the base sequential model as the reranker.





\subsection{KG-enhanced Personalized Information Extraction}
Existing methods 
primarily rely on item interaction sequences to infer user intent. We leverage heterogeneous KG to infer user personas that reflect stable, long-term user preferences derived from semantic and relational information.


\subsubsection{Heterogeneous KG Construction and Initialization}
We construct a hybrid knowledge graph (KG) that integrates collaborative signals (without timestamps), item semantics, and external knowledge. We utilize Amazon-KG \cite{wang2024amazon}\footnote{\label{amazonkg}\url{https://github.com/WangYuhan-0520/Amazon-KG-v2.0-dataset}} which enriches Amazon interaction data with structured information derived from DBPedia. This hybrid KG encodes time-independent user preferences, complementing session-based interaction sequences that primarily capture short-term intent.

\paragraph{LLM-based Node and Edge Initialization 
.} To incorporate semantic information from item metadata, we initialize node and relation embeddings using an LLM. This initialization injects semantic knowledge derived from textual metadata into the heterogeneous graph, aligning heterogeneous node and relation types within a unified embedding space. User nodes remain free parameters to be discovered by graph‑level self‑supervision.
We initialize the KG with text-derived semantics as follows:
\begin{itemize}
    \item \textbf{Item nodes:} For each item $i$, we encode textual fields (e.g. title and description) and encode them with an \textbf{LLM} to obtain a semantic embedding $\mathbf{e} ^ {\text{text}}_i \in \mathbb{R}^{d_t}$.
    \item \textbf{Meta data nodes:} For each attribute $ \mathbf{a} \in \mathcal{A}$, we initialize node features $\mathbf{x}_a$ using LLM-based encodings of attribute descriptions.
    \item \textbf{Edges / relations:} If relation types $r$ (or edge texts) have textual descriptors, we encode them with an LLM to obtain $\mathbf{e}^{\text{text}}_r$.
    \item \textbf{User nodes:} Since sessions are anonymous and users lack textual metadata, user node embeddings are randomly initialized: $\mathbf{x}_u \!\sim\! \mathcal{N}(0, \sigma^2\mathbf{I})$.
\end{itemize}
These embeddings are later refined through unsupervised graph representation learning.

\subsubsection{Unsupervised User Representation Learning via Heterogeneous Deep Graph Infomax} 

To capture rich and semantically grounded user preferences from the heterogeneous knowledge graph, we employ an unsupervised representation learning framework based on Heterogeneous Deep Graph Infomax (HDGI) \cite{ren2019heterogeneous}.
\paragraph{Heterogeneous Graph Encoding.}
Following \cite{ren2019heterogeneous}, the heterogeneous KG consists of multiple node types (e.g., users, items, attributes, entities) and relation types. HDGI utilizes meta-path structures to capture semantic connectivity patterns across heterogeneous nodes. For a given relation type $r$, the neighborhood aggregation is performed through a graph convolution module specialized for heterogeneous graphs. For each node $v$, the encoder produces a local representation:
\[
    \mathbf{h}_v = f_\theta\big(v, \mathcal{N}_r(v)\big),
\]
where $\mathcal{N}_r(v)$ denotes the type-specific neighborhood under relation $r$. A semantic-level attention mechanism is used to fuse meta-path-based embeddings, enabling the model to preserve heterogeneous relational semantics \cite{ren2019heterogeneous}.

\paragraph{Local-Global Mutual Information Maximization.}
The core Infomax objective is motivated by DGI \cite{velivckovic2018deep} principle, which maximizes mutual information between node-level features and a global summary of the graph. HDGI extends this idea to the heterogeneous setting. The global summary vector is computed as:
\[
    \mathbf{s} = \sigma\left( \frac{1}{|V|} \sum_{v \in V} \mathbf{h}_v \right).
\]
To construct the contrastive objective, HDGI generates corrupted node embeddings $\tilde{\mathbf{h}}_v$ by shuffling node features or perturbing structural connections. A discriminator $D_\phi$ then distinguishes between positive (real) and negative (corrupted) pairs:
\[
    \mathcal{L}_{\text{HDGI}} = - \sum_{v \in V} \left[ 
        \log D_\phi(\mathbf{h}_v, \mathbf{s}) 
        + 
        \log\!\left(1 - D_\phi(\tilde{\mathbf{h}}_v, \mathbf{s})\right)
    \right].
\]
This contrastive Infomax objective encourages the encoder to extract globally consistent, semantically aligned node embeddings.

\paragraph{User Persona Embedding Extraction.}
After HDGI training, we directly extract the embeddings of the user nodes from the shared heterogeneous representation space to represent user personas. Because HDGI incorporates meta-path semantics, graph convolution aggregation, and mutual-information-based contrastive learning, the resulting embeddings encode multi-hop, multi-relational preference signals. 





%



\subsection{KG-Enhanced Personalized Information Utilization}
The retrieved candidate items are further refined using the base sequential model as the reranker to focus more on short-term intent. Given a session sequence the sequential model encodes the session context and computes relevance scores for candidate items. The model outputs relevance scores over the candidate set and the top-K items are returned as the final recommendations. 


\paragraph{}
By combining KG-derived persona embeddings with SBRS, the proposed framework enables the retriever to utilize both short-term behavioral signals and long-term preference information encoded in the knowledge graph. This design improves personalization under the anonymous session assumption.
\section{Experiments}

\subsection{Datasets}
    We evaluate our framework on two publicly available benchmark datasets from the Amazon Reviews \footnote{\url{https://cseweb.ucsd.edu/~jmcauley/datasets/amazon_v2/}} from amazon product recommendation domain. Amazon Books and Amazon Movies \& TV. These datasets contain user-item interaction sequences and rich item metadata such as titles, categories, and descriptions. For each dataset, we retained users and items with at least 5 interactions and ordered them chronologically. The statistics of the dataset is provided in Table \ref{tab:stats}. 
    
    \subsubsection{Construction of KG} 
\begin{figure}
    \centering
    \includegraphics[width=1.1\textwidth]{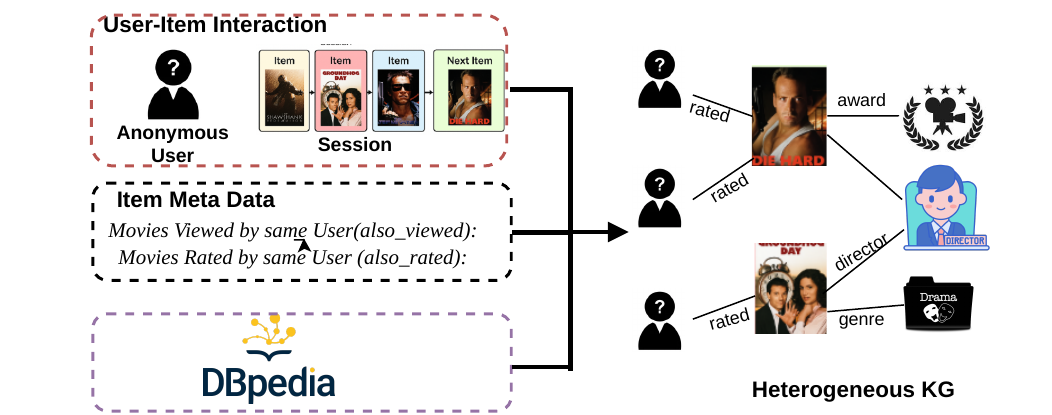}

    \caption{Construction of KG}
    \label{fig:my_label}

\end{figure}

    To construct the heterogeneous knowledge graph, we utilize Amazon-KG \footnote{\url{https://github.com/WangYuhan-0520/Amazon-KG-v2.0-dataset}}, which enriches interaction data with structured item attributes and entity relations derived from DBpedia as shown in Fig: \ref{fig:my_label}. This hybrid knowledge graph enables the model to capture semantic and relational signals beyond raw interaction sequences. 
    The heterogeneous knowledge graph contains multiple node types, including \textit{User}, \textit{Movie}, \textit{Person}, \textit{Subject}, \textit{Genre}, \textit{Award}, \textit{OpeningTheme}, \textit{LiteraryGenre}, and \textit{Category} for the Amazon Movies \& TV dataset, and \textit{User}, \textit{Book}, \textit{Person}, \textit{Subject}, \textit{Genre}, \textit{LiteraryGenre}, \textit{Series}, and \textit{NotableWork} for the Amazon Books dataset. The heterogeneous knowledge graph contains multiple relation types, including 
\textit{(Movie, subject\_is, Subject)}, 
\textit{(Movie, directed\_by, Person)}, 
\textit{(Movie, musiccomposer\_is, Person)}, 
\textit{(Movie, starring, Person)}, 
\textit{(Movie, writer\_is, Person)}, 
\textit{(Movie, composer\_is, Person)}, 
\textit{(Movie, executiveProducer\_is, Person)}, 
\textit{(Movie, genre\_is, Genre)}, 
\textit{(Movie, got\_award, Award)}, 
\textit{(Movie, openingTheme, OpeningTheme)}, 
\textit{(Movie, previousWork, Movie)}, 
\textit{(Movie, subsequentWork, Movie)}, 
\textit{(Movie, creator, Person)}, 
\textit{(Movie, presenter, Person)}, 
\textit{(Movie, have\_artist, Person)}, 
\textit{(Movie, literaryGenre\_is, LiteraryGenre)}, 
\textit{(Movie, also\_view\_product, Movie)}, 
\textit{(Movie, category, Movie)}, 
\textit{(Movie, brand, Person)}, 
\textit{(Movie, also\_buy\_product, Movie)}, 
and \textit{(User, rated, Movie)} for the Amazon Movies \& TV dataset, and 
\textit{(Book, subject\_is, Subject)}, 
\textit{(Book, author, Person)}, 
\textit{(Book, musiccomposer\_is, Person)}, 
\textit{(Book, director, Person)}, 
\textit{(Book, writer\_is, Person)}, 
\textit{(Book, composer\_is, Person)}, 
\textit{(Book, executiveProducer\_is, Person)}, 
\textit{(Book, genre\_is, Genre)}, 
\textit{(Book, series, Series)}, 
\textit{(Book, producer, Person)}, 
\textit{(Book, previousWork, Book)}, 
\textit{(Book, subsequentWork, Book)}, 
\textit{(Book, creator, Person)}, 
\textit{(Book, starring, Person)}, 
\textit{(Book, notableWork, NotableWork)}, 
\textit{(Book, literaryGenre\_is, LiteraryGenre)}, 
and \textit{(User, rated, Book)} for the Amazon Books dataset. The statistics of number of entities and relations is provided in Table \ref{tab:stats}.

\begin{table}[t]
\caption{Statistics of the datasets used for experiments.}
\label{tab:stats}
\begin{tabular}{lcccccc}
\hline
{ \textbf{Datasets}} &{{ \textbf{\#users}}} & { \textbf{\#items}} & { \textbf{\#interactions}} & {\textbf{\#entities}} & {\textbf{\#triples}}\\ \hline
{ Amazon-Movies \& TV}     & {{10,849 }}          & {3,672 }          & {87,376 }           & { 45,528} & {20,9630}\\ \hline
{ Amazon-Books}     & {{1,624 }}           & {1,345 }           & { 10,947}            & { 8,660} & {22,830} \\ \hline
\end{tabular}
\end{table}

\subsection{Experimental setup}
To evaluate the effectiveness of different user preference representations, we compare different strategies for incorporating user preference representations in the retrieval module. Specifically, we evaluate four configurations of user preferences: 
\begin{itemize}
    \item \textbf{None}: No user preference representation is used.
    \item \textbf{Random}: User preference representations are randomly initialized, and learned during training.
    \item \textbf{LLM}: User preference representations are generated using an LLM based on textual item signals. The prompt is provided in Table \ref{tab:prompts}
    \item \textbf{KG-LLM}: User preference representations are learned from the heterogeneous knowledge graph using the proposed HDGI-based representation learning approach. The item node embeddings with textual descriptions are intialized using an LLM.
\end{itemize}
Item representations are obtained by encoding item's textual descriptions using an LLM. The description of the item is generated from item's title using an LLM. Prompt is provided in Table \ref{tab:prompts}. We use Qwen-3-8B \cite{qwen3technicalreport}\footnote{\url{https://huggingface.co/Qwen/Qwen3-8B}} model wherever an LLM call is required for text generation, and Qwen3-Embedding-8B\footnote{\url{https://huggingface.co/Qwen/Qwen3-Embedding-8B}} for text embeddings. We report standard top-$K$ recommendation metrics including Hit Rate (HR@K), Normalized Discounted Cumulative Gain (NDCG@K), and Mean Reciprocal Rank (MRR). We perform reranking using base sequential model (SASRec). 

\begin{table}[t]
\centering
\caption{Prompts used for user persona generation.}
\label{tab:prompts}
\begin{tabular}{p{3cm}|p{9cm}}
\hline
\textbf{Prompt Type} & \textbf{Prompt} \\
\hline

User Preference 
&
Based on the titles and features of the items the user has interacted with in chronological order, summarize the user's preferences directly as concise and precise keywords, separated by commas. No additional commentary; exclude lines like "Here is a summary". Strictly output only the user's preferences.

\textbf{Example output:} Drama Movie, Comedy, Action, Character Investment

\textcolor{blue}{\textit{\textbf{<History>}}} \\
\hline

Item Description & Based on the Movies or TV name and your general knowledge, provide an objective description of each movie or TV. Ensure the response is concise and informative. No Additional Commentary. Strictly output only response. 

    \textbf{Example}:
    Movie and TV name: Dazed and Confused
    Response:
    It was bongs and bell bottoms, polyester and puka shells, macrame and mood rings. We rocked and rolled, we were jaded and innocent, but most of all, we were Dazed and Confused. With hilarious and touching honesty, this critically acclaimed comedy explores the last day of school - and one rowdy night - in the lives of a group of high school students in late May, 1976. Accompanied by the music of Aerosmith, Alice Cooper, Foghat and more, a superb ensemble cast delivers "the most slyly funny and dead-on portrait of American teenage life ever made." (Entertainment Weekly)
    
    Now, analyze the following Movie and TV name:
    \textcolor{blue}{\textit{\textbf{<Movie and TV Name>}}} 
    
    \textbf{Response:} \\
\hline

\end{tabular}

\end{table}

\subsubsection{Hyperparameter Setup}
Our base sequential model is SASRec \cite{Kang2018SASRec}. We use validation data for hyperparameter selection using NDCG@10 as the performance metric for all approaches. We tuned the architectural parameters over Transformer depth $L \in \{2,3,4\}$, hidden dimensionality $d \in \{64,128,256\}$, number of attention heads $\{2,4\}$, and dropout rates $\{0.1,0.2,0.3,0.5\}$. Optimization settings were explored over learning rates $\{1\!\times\!10^{-4},\, 3\!\times\!10^{-4},\, 5\!\times\!10^{-4},\, 1\!\times\!10^{-3}\}$, and batch sizes $\{128, 256, 512, 1024\}$. The results shown in Table~\ref{tab:retrieval_results} correspond to the configuration with $d=64$, dropout $=0.1$, learning rate $=1e-3$, and $L=2$. All models were trained using the Adam optimizer. 

\subsection{Results and Discussion}
Table \ref{tab:retrieval_results} compares the performance of different user representation strategies when used in the retriever (data-driven recommendation with user embeddings) module. The results demonstrate that incorporating knowledge-graph-derived user personas consistently improves recommendation performance across both datasets. The inclusion of KG-based user preference embeddings leads to higher HR@100 across both datasets, indicating that the correct item is more frequently included in the retrieved candidate set. This improvement suggests that the heterogeneous KG captures stable long-term user preferences through relational signals, enabling the retriever to generate more relevant candidate items. For the Amazon Books dataset, improvements in HR@10 are less pronounced due to the high sparsity of interaction data. In such cases, KG-based user representations primarily improve candidate recall rather than top-rank precision. The heterogeneous knowledge graph introduces additional relational signals that allow the retriever to identify more relevant items, including long-tail items connected through semantic relations. However, these items may appear lower in the ranked list, making the improvement more visible when the candidate cutoff increases to k=100. After candidate generation, we apply reranking using base SASRec without explicit user embeddings and with randomly initialized item embeddings. Although the reranker does not directly use KG-based persona information, it improves MRR and NDCG, which measure the ranking quality within the candidate set. This improvement occurs because SASRec models sequential dependencies within the session, allowing the model to better order candidate items according to the user's short-term intent.


\begin{table}[t]
\centering
\caption{\centering Performance comparison of user representation strategies used in the retriever module}
\resizebox{\linewidth}{!}{
\begin{tabular}{c|c|ccc|ccc}
\hline
\multirow{2}{*}{\begin{tabular}{c}\textbf{User}\\\textbf{Embedding}\end{tabular}} 
& \multirow{2}{*}{\begin{tabular}{c}\textbf{Item}\\\textbf{Embedding}\end{tabular}} 
& \multicolumn{3}{c|}{\textbf{Amazon Movies \& TV}} 
& \multicolumn{3}{c}{\textbf{Amazon Books}} \\

 & & \textbf{HR} & \textbf{NDCG} & \textbf{MRR} & \textbf{HR} & \textbf{NDCG} & \textbf{MRR} \\
\hline

\multicolumn{8}{c}{\bm{$k=10$}}\\
\hline

None & Random & 0.1302 &0.0752 &0.0585 
  & 0.1693 & 0.1109  & 0.093  \\

None & Description (LLM) &0.1314 &0.0774 &0.061  &0.1791  & 0.1075 & 0.0857 \\

Random & Description (LLM) 
& 0.1116 &0.0624 &0.0473
& 0.1711 & 0.1005 & 0.0788 \\

LLM & Description (LLM) 
& 0.1171 &0.0648 &0.049
& 0.1785 & 0.1077 & 0.086 \\

KG+LLM & Description (LLM) 
& \textbf{0.1327} &0.0764 &0.0593
& 0.1773 & 0.1101 & 0.0894 \\
\hline
\multicolumn{8}{c}{\textbf{Reranking}}\\
\hline
None & Random & \textbf{0.1327} & \textbf{0.776} & \textbf{0.0609} & \textbf{0.1773} & \textbf{0.1123} &	\textbf{0.0925} \\

\hline
\hline
\multicolumn{8}{c}{\bm{$k=100$}}\\
\hline

None & Random & 0.3665 &	0.1220 &	0.0666
  & 0.4254 & 0.1607 &	0.1012   \\

None & Description (LLM) & 0.3639 &	0.1235 &0.069 & 0.427 & 0.1559&	0.0939   \\

Random & Description (LLM) 
&  0.3507 &	0.1097 &0.0555
& 0.4347 & 0.1525 &	0.0877   \\

LLM & Description (LLM) 
& 0.3527 &0.111 &	0.0568
& 0.4384 &  0.1588 &	0.0948 \\

KG+LLM & Description (LLM) 
& \textbf{0.3697} & \textbf{0.1229} &0.0672
& \textbf{0.4427}  &  \textbf{0.1622} &	0.0982 \\

\hline
\multicolumn{8}{c}{\textbf{Reranking}}\\
\hline
None & Random & \textbf{0.3697} & \textbf{0.1236} & \textbf{0.0677} & \textbf{0.4427} & \textbf{0.1641} &	\textbf{0.1016} \\
\hline
\end{tabular}
}
\label{tab:retrieval_results}
\end{table}

\section{Conclusion}

In this work, we proposed a persona-driven framework for session-based recommendation that integrates heterogeneous knowledge graphs and LLMs to bridge short-term session intent and stable long-term user preferences. Our approach learns KG-grounded user personas in an unsupervised manner using Heterogeneous Deep Graph Infomax and combines them with LLM-derived item semantic embeddings for improved candidate retrieval and ranking. Experiments on Amazon Books and Amazon Movies \& TV show that incorporating KG-derived user personas significantly improves recommendation performance compared to sequential baselines relying only on session history. These results demonstrate the effectiveness of combining structured relational knowledge with semantic representations for personalized recommendation in sparse and anonymous session settings. 


\end{document}